\documentclass[final]{elsarticle}

\usepackage{amsmath}
\usepackage{amssymb}
\usepackage{amsthm}
\usepackage{graphicx}
\usepackage{url}

\newcommand{\D}{\displaystyle}
\font\SSI=cmssi12
\newcommand{\Z}{\mathop{\hbox{\SSI Z\kern-.4em Z}}\nolimits}

\newcommand{\lcm}{\mathrm{lcm}}

\journal{}

\begin{document}

\begin{frontmatter}

\title{Heterogeneous, Weakly Coupled Map Lattices}

\author[1]{M$^\mathrm{a}$ Dolores Sotelo Herrera}
\author[1]{Jes\'{u}s San Mart\'{i}n}
\ead{jsm@dfmf.uned.es}
\address[1]{Departamento de Matem\'atica Aplicada, E.T.S.I.D.I., Universidad Polit\'ecnica de Madrid, Madrid, Spain}

\author[2]{Mason A. Porter}
\address[2]{Oxford Centre for Industrial and Applied Mathematics, Mathematical Institute, University of Oxford, Oxford, United Kingdom}

\begin{abstract}

Coupled map lattices (CMLs) are often used to study emergent phenomena in nature. It is typically assumed (unrealistically) that each component is described by the same map, and it is important to relax this assumption. In this paper, we characterize periodic orbits and the laminar regime of type-I intermittency in \emph{heterogeneous} weakly coupled map lattices (HWCMLs). We show that the period of a cycle in an HWCML is preserved for arbitrarily small coupling strengths even when an associated uncoupled oscillator would experience a period-doubling cascade. Our results characterize periodic orbits both near and far from saddle--node bifurcations, and we thereby provide a key step for examining the bifurcation structure of heterogeneous CMLs.
\end{abstract}


%HIGHLIGHTS
% Analytical solutions for heterogeneous weakly coupled map lattices
% Periodic orbits far from and near saddle--node bifurcations
% Analytical expression of laminar regime of a type-I intermittency

\end{frontmatter}
%%%%%%%%

\section{Introduction}

Numerous phenomena in nature --- such as human waves in stadiums \cite{farkas03} and flocks of 
seagulls \cite{sumpter10} --- result from the interaction of many individual elements, and they can exhibit fascinating emergent dynamics that cannot arise in individual or even small numbers of components \cite{nino}. In practice, however, a key assumption in most such studies is that each component is described by the same dynamical system.  However, systems with heterogeneous elements are much more common than homogeneous systems.  For example, a set of interacting cars on a highway that treats all cars as the same ignores different types of cars (e.g., their manufacturer, their age, different levels of intoxication among the drivers, etc.), and a dynamical system that governs the behavior of different cars could include different parameter values or even different functional forms entirely for different cars. Additionally, one needs to use different functional forms to address phenomena such as interactions among cars, traffic lights, and police officers. Unfortunately, because little is known about heterogeneous interacting systems \cite{Coca,Pavlov}, the assumption of homogeneity is an important simplification that allows scholars to apply a plethora of analytical tools. Nevertheless, it is important to depart from the usual assumption of homogeneity and examine coupled dynamical systems with heterogeneous components. 

The study of coupled map lattices (CMLs) \cite{librokaneko,kaneko25} is one important way to study the emergent phenomena (e.g., cooperation, synchronization, and more) that can occur in interacting systems. CMLs have been used to model systems in numerous fields that range from physics and chemistry to sociology, economics, and computer science \cite{kaneko25,Special1992,Special1997,Tang2010,Wang2014}.  In a CML, each component is a discrete dynamical system (i.e., a map). There are a wealth of both theoretical and computational studies of homogeneous CMLs \cite{librokaneko,kaneko25,Kaneko89,He1996,Franceschini02,Cherati07,Jakobsen08,Herrera09,Xu10}, in which the interacting elements are each governed by the same map. Such investigations have yielded insights on a wide variety of phenomena. As we mentioned above, the assumption of homogeneity is a major simplification that often is not justifiable. Therefore, we focus on heterogeneous CMLs, in which the interacting elements are governed by different maps or by the same map with different parameter values. The temporal evolution of a heterogeneous coupled map lattice (CML) with $p$ components is given by
 \begin{equation}\label{equ1}
	X_i(n+1)=f_{R_i}(X_{i}(n)) + \varepsilon \D \sum_{\begin{subarray}{c}h=1\\h\not=i \end{subarray}}^p f_{R_h}(X_{h}(n))\,, \qquad i\in \{1,\dots,p\}\,,
\end{equation}
where $X_i(n)$ represents the state of the entity at instant $n$ at position $i$ of a lattice and $\varepsilon > 0$ weights the coupling between these entities.  We consider entities in the form of oscillators, where the $i$th oscillator evolves according to the map 
\begin{equation}\label{equb}
	X_i(n+1)=f_{R_i}(X_{i}(n))\,, \qquad i\in\{1,\dots,p\}\,, 
\end{equation}
where the $f_{R_i}$ are, in general, different functions that depend on a parameter $R_i$ (where $i \in \{1,\dots,p\}$).  We assume that each $f_{R_i}$ is a $C^2$ unimodal function that depends continuously on the parameter $R_i$ with a critical point C at $R_i$. As usual, $f^m$ means that $f$ is composed with itself $m$ times. If an uncoupled oscillator $X_i(n)$ takes the value $x_{i,n}$, then the evolution of this value under the map is $x_{i,n+1} = f_{R_i}(x_{i,n})$.

In this paper, we examine heterogeneous, weakly coupled map lattices (HWCMLs). Weakly coupled systems can exhibit phenomena (e.g., phase separation because of additive noise \cite{Angelini01}) that do not arise in strongly coupled systems, and one can even use weak coupling along with noise to fully synchronize nonidentical oscillators \cite{yiming}. Thus, it is important to examine HWCMLs, which are amenable to perturbative approaches.  In our paper, we characterize periodic orbits both far away from and near saddle--node (SN) bifurcations.  Understanding periodic orbits is interesting by itself and is also crucial for achieving an understanding of more complicated dynamics (such as chaos) \cite{scholarpedPO,chaosbook}. We then characterize the laminar regime of type-I intermittency in our HWCMLs. Finally, we summarize our results and briefly comment on applications.

%%%%%%%%

\section{Theoretical Results}\label{theory}

Before discussing our results, we need to define some notation. Let $x_{i,n|R_i}$ denote the points in a periodic orbit of the $i$th uncoupled oscillator with control parameter $R_i$. The parameter value $r_i$ is a bifurcation value of $R_i$ for the $i$th map, so $x_{i,n|r_i}$ denotes the points in a periodic orbit at this parameter value.

Suppose that $R_i = r_i + \varepsilon^\alpha$, where $\varepsilon$ is the same as in the coupling term of the CML (\ref{equ1}) and $\alpha \in (0,\infty)$ is a constant. We seek to derive results that are valid at size $O(\varepsilon)$. We need to consider the following situations:
\begin{enumerate}
\item[$\alpha < 1$] In this case, when we expand to size $O(\varepsilon)$, the coupling term does not contribute at all. Therefore, the oscillators in (\ref{equ1}) behave as if they were uncoupled at this order of the expansion.
\item[$\alpha > 1$] In this case, the coupling term controls the $\varepsilon$ bifurcation terms. Thus, to size $O(\varepsilon)$, we cannot study the behavior of the bifurcation.
\item[$\alpha = 1$] In this case, we are considering a perturbation of the same size as the coupling term, and we can simultaneously study the coupling and the bifurcation analytically.
\end{enumerate}

To study orbits close to bifurcation points, we thus let $R_i = r_i + \varepsilon$, where $\varepsilon$ is the same as in the coupling term of the CML (\ref{equ1}). In our numerical simulations (see Section \ref{sec:numerical}), we will also briefly indicate the effects of considering $\alpha \neq 1$ (see Section \ref{alp}).

%%%%%

\subsection{Study of the CML Far from and Close to Saddle--Node Bifurcations}\label{sub:PeriodicOrbits}

In this section, we examine heterogeneous CMLs in which the uncoupled oscillators have periodic orbits either far from or near SN bifurcations. As periodic orbits exhibit different dynamics from each other depending on whether they are near or far from SN bifurcations \cite{Sotelo10,Sotelo12}, it is important to distinguish between these two situations.

A period-$m$ SN orbit is a periodic orbit that is composed of $m$ ``SN points'' of the composite map $f_{r_i}^m$. Each of these $m$ SN points is a fixed point of $f_{r_i}^m$ at which $f_{r_i}^m$ undergoes an SN bifurcation. Period-$m$ SN orbits play an important role in a map's bifurcation structure, because they occur at the beginning of periodic windows in bifurcation diagrams. Studying them is thus an important step towards examining the general bifurcation structure of a map.

\begin{figure}[!hhh]
	\includegraphics[width=0.5\textwidth]{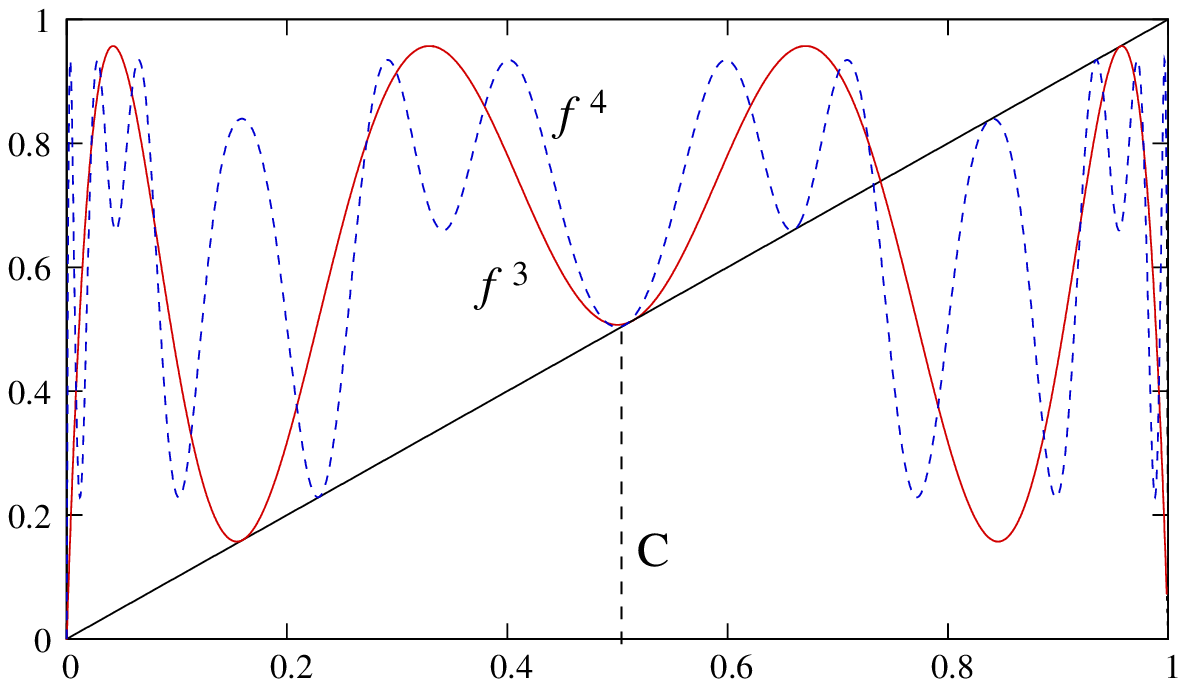}\includegraphics[width=0.5\textwidth]{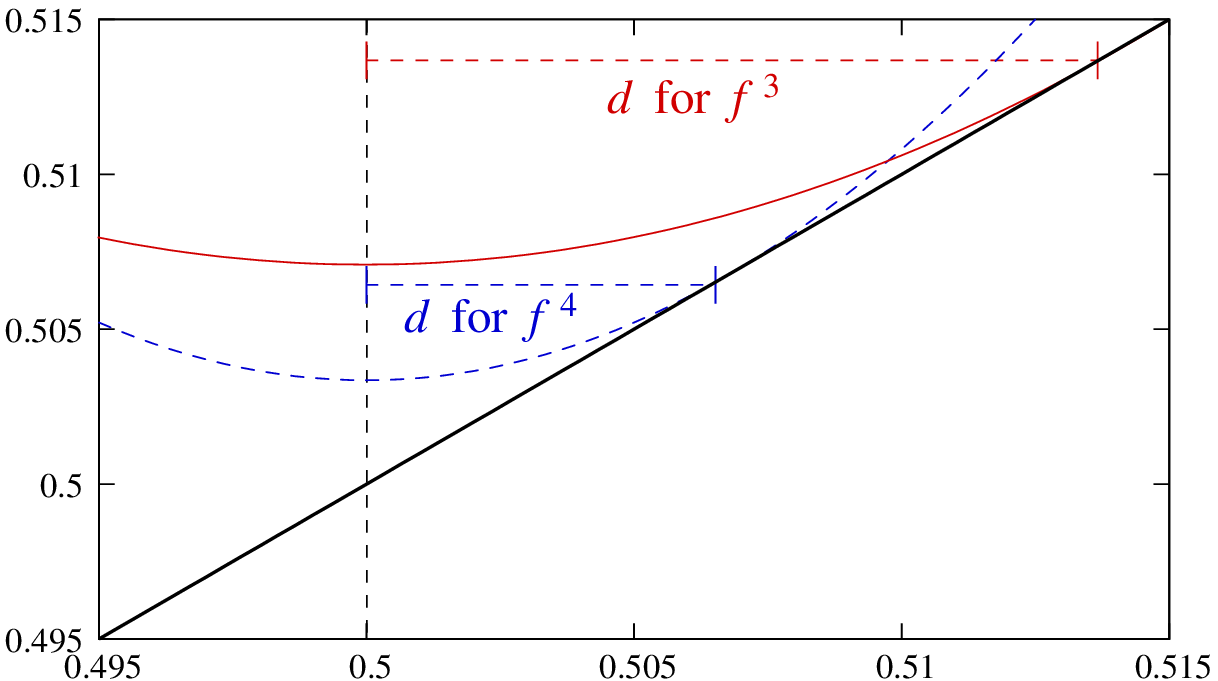}
	\caption{\label{fig:SN}
             {\bf Left: The maps $f^3$ and $f^4$ (where $f$ is the logistic map) and the fixed points at which SN bifurcations occur}. Observe that there are SN points far away from the critical point C. Because $4 > 3$, the extremum of $f^4$ near the critical point C is narrower than the extremum of $f^3$ near C. 
	{\bf Right: Magnification of the extrema near the critical point C}. We show the distances $d$ between the SN point and the critical point for both $f^3$ and $f^4$. Observe that the distance between this pair of points decreases as the period increases.
	}
\end{figure}

When  $f_{r_i}^m$ undergoes an SN bifurcation, the map $f_{r_i}$ has two properties that we highlight. Let $\{x_{i,1|r_i},x_{i,2|r_i}\dots ,x_{i,m|r_i}\}$ be an period-$m$ SN orbit. It then follows that
\begin{enumerate}
\item \begin{equation}
	\nonumber  \frac{\partial f^{m}_{r_i}}{\partial x} (x_{i,j|r_i}) = 1=\D\prod_{k=j}^{j+m-1}\frac{\partial f_{r_i}}{\partial x}(x_{i,k|r_i})
\end{equation}
Consequently, orbits that are near an SN orbit satisfy
\begin{equation}\label{12A}
	1 - \prod_{k=j}^{j+m-1} \D\frac{\partial f_{R_i}}{\partial x} (x_{i,k|R_i}) = o\D(1)\,.
\end{equation}
By contrast, if
\begin{equation}\label{12B}
	1 - \prod_{k=j}^{j+m-1} \D\frac{\partial f_{R_i}}{\partial x} (x_{i,k|R_i}) = O\D(1)\,.
\end{equation}
we say that an orbit is ``far from'' a SN orbit. 
\item{Because $f_{r_i}$ has a critical point at C, so does $f_{r_i}^m$. Suppose that $x_{i,n|r_i}$ is the point of the SN orbit that is closest to C. As $\D\frac{\partial f_{r_i}}{\partial x}(\mbox{C})=0$, for sufficiently large periods, we can find SN orbits with  
arbitrarily small $\left|\D\frac{\partial f_{r_i}}{\partial x}(x_{i,n|r_i})\right|$ (see Fig.~\ref{fig:SN}), and in particular we can find examples where $\left|\D\frac{\partial f_{r_i}}{\partial x}(x_{i,n|r_i})\right| < \varepsilon$. We use the term \emph{small-derivative} SN orbits for such orbits. Additionally, a \emph{small-derivative} SN orbit includes points that are not close to the critical point $\mathrm{C}$, so that $\frac{\partial f}{\partial x}(x_i) = O(1)$ in general, and the associated terms cannot be neglected.
}
\end{enumerate}

%%%%

The overall 
bifurcation pattern in a typical unimodal map of the interval is topologically equivalent to the bifurcation pattern in any other typical unimodal map of the interval \cite{gucken}, so it is sensible to focus on a particular such map. The standard choice for such a map is the logistic map. Orbits of any period occur in the logistic map, which contains infinitely many \emph{small-derivative} SN orbits. In particular, such orbits include the period-$q$ SN orbits from which supercycles with symbol sequences $\mbox{CR}\mbox{L}^{q-2}$ originate.\footnote{Recall that a ``supercycle'' is a periodic orbit that includes C; if its period is $q$ (i.e., if $f^q_R(C)=C$ for some parameter value $R$), then $\D\frac{\partial f^q_{R}}{\partial x}(x_k)=0$ for all points $x_k$ in the supercycle.}
Given this fact and the broad applicability of results for the logistic map, we note that our results are relevant in numerous situations.

%%%%%

\subsection*{Lemma 1}

Let $|\varepsilon| < 1$ in the CML (\ref{equ1}), and suppose that the map $f_{R_i}^{q_i}$ has an SN bifurcation at $R_i = r_i$, such that the associated SN orbit of $f_{r_i}$ is a {small-derivative} SN orbit. Additionally, suppose that $R_i=r_i + \varepsilon$ for $i\in\{1,\dots,s\}$, but that $R_i$ for $i\in\{s+1,\dots,p\}$ are far away from $r_i$. 
Consider the following initial conditions:
\begin{itemize}
\item For $i\in\{1,\dots,s\}$, let $X_i (n) = x_{ i,n|r_i} + \varepsilon A_{i,n}+O(\varepsilon^2)$, where $x_{ i,n|r_i}$ is the point of the SN orbit closest to the critical point C of $f_{R_i}$ at $R_i=r_i$.
\item For $i\in\{s+1,\dots,p\}$, let $X_i (n) = x_{ i,n|R_i} + \varepsilon A_{i,n}+O(\varepsilon^2)$.
\end{itemize}
The temporal evolution of the CML (\ref{equ1}) is then given by
\begin{enumerate}
\item{For $i\in\{1,\dots,s\}$, 
 \begin{align}\label{oneone}
X_i(n+1) &= x_{i,n+1|r_i}+\varepsilon \left(\D \frac{\partial f_{r_i}}{\partial r}(x_{i,n|r_i}) +\D \sum_{\substack{
       h=1\\h\not=i}}^s x_{h,n+1|r_h}+\sum_{\substack{
       h=s+1}}^p x_{h,n+1|R_h}  \right)+O(\varepsilon^2)\,, \quad m = 1\,,
 \end{align}
%and for $m \geq 2$
\begin{align}\label{one}
X_i(n+m) &= x_{i,n+m|r_i}+\left[ \D\frac{\partial f_{r_i}}{\partial r}(x_{i,n+m-1|r_i})+\D\sum_{k=n}^{n+m-2} \frac{\partial f_{r_i}}{\partial r}(x_{i,k|r_i})\prod_{l=k+1}^{n+m-1} \frac{\partial f_{r_i}}{\partial x}(x_{i,l|r_i})\right. \notag \\
&\qquad+\D\sum_{k=n+1}^{n+m-1} \left(\left( \sum_{\substack{
            h=1\\h\not=i}}^s x_{h,k|r_h}+\sum_{\substack{
            h=s+1}}^p x_{h,k|R_h} \right) \prod_{l=k}^{n+m-1} \frac{\partial f_{r_i}}{\partial x}(x_{i,l|r_i}) \right)\notag \\
            &\qquad \left.+\sum_{\substack{
            h=1\\h\not=i}}^s x_{h,n+m|r_h}+\sum_{\substack{  h=s+1}}^p x_{h,n+m|R_h} \right]\varepsilon+O(\varepsilon^2)\,, \quad m \in \{2,\dots,q\}.
\end{align}
}
\item{For $i\in\{s+1,\dots,p\}$, 
\begin{equation}\label{twoone}
	X_i(n+1) =x_{i,n+1|R_i}+\varepsilon \left(\D \frac{\partial f_{R_i}}{\partial x}(x_{i,n|R_i}) A_{i,n}+\D \sum_{\substack{
       h=1}}^s x_{h,n+1|r_h}+\sum_{\substack{
       h=s+1\\h\not=i}}^p x_{h,n+1|R_h}\right)+O(\varepsilon^2)\,, \quad m = 1\,,
\end{equation}
%and for $m \geq 2$
\begin{align}\label{two}
\nonumber X_i(n+m)&=x_{i,n+m|R_i}+\left[ \D\prod_{k=n}^{n+m-1} \frac{\partial f_{R_i}}{\partial x}(x_{i,k|R_i})A_{i,n}+\D\sum_{k=n+1}^{n+m-1} \left( \left(\D \sum_{\substack{h=1}}^s  x_{h,k|r_h}\sum_{\substack{h=s+1\\h\not=i}}^p + x_{h,k|R_h}\right)\right.\right.\\
                    &\qquad\left. \left. \D\prod_{l=k}^{n+m-1} \frac{\partial f_{R_i}}{\partial x}(x_{i,l|R_i})\right)+\D\sum_{\substack{h=1}}^s x_{h,n+m|r_h}+\sum_{\substack{h=s+1\\h\not=i}}^p x_{h,n+m|R_h}\right]\varepsilon+O(\varepsilon^2)\,, \quad m \in \{2,\dots,q\}.
\end{align}
}
\end{enumerate}

%%%%%%

\subsection*{Proof of Lemma 1}

We proceed by induction. Substitute $X_i(n)=x_{i,n|r_i}+\varepsilon A_{i,n}+O(\varepsilon^2)$ and $X_i(n)=x_{i,n|R_i}+\varepsilon A_{i,n}+O(\varepsilon^2)$ into equation (\ref{equ1}) and expand in powers of  $\varepsilon$. Note that we need to consider $i\in\{1,\dots, s\}$ and $i\in \{s+1,\dots, p\}$ separately.

\begin{enumerate}
\item
We initiate the iteration at the point $x_{i,n|r_i}$ of the SN orbit closest to the critical point C. Because we have a small-derivative SN orbit, $\frac{\partial f_{r_i}}{\partial x}(x_{i,n|r_i})$ is arbitrarily small, although this is not true for other points in the SN orbit.

\vspace{5mm}
For $ i\in\{1,\dots, s\}$, we have
 \begin{align}\label{eq12}                   
X_i(n+1) &= f_{r_i}(x_{i,n|r_i})+\varepsilon\frac{\partial f_{r_i}}{\partial x}(x_{i,n|r_i})A_{i,n}+\varepsilon \frac{\partial f_{r_i}}{\partial r}(x_{i,n|r_i}) \notag \\ &
+\varepsilon \D \sum_{\substack{
       h=1\\h\not=i}}^s f_{r_h}(x_{h,n|r_h}+O(\varepsilon))+ \sum_{\substack{
       h=s+1}}^p f_{R_h}(x_{h,n|R_h}+O(\varepsilon))+O(\varepsilon^2) \notag \\ &
       	= x_{i,n+1|r_i}+\varepsilon \left(\D \frac{\partial f_{r_i}}{\partial r}(x_{i,n|r_i}) +\D \sum_{\substack{
       h=1\\h\not=i}}^s x_{h,n+1|r_h}+\sum_{\substack{
       h=s+1}}^p x_{h,n+1|R_h}  \right)+O(\varepsilon^2)\,,
 \end{align}
In the last step, we have neglected terms that contain $\varepsilon \D\frac{\partial f_{r_i}}{\partial x}(x_{i,n|r_i})$ because $\D\frac{\partial f_{r_i}}{\partial x}(x_{i,n|r_i})$ is arbitrarily small.

\item{For $i\in \{s+1\,,\dots, p\}$, we have
\begin{equation}\label{eq121}                   
	X_i(n+1) =x_{i,n+1|R_i}+\varepsilon \left(\D \frac{\partial f_{R_i}}{\partial x}(x_{i,n|R_i}) A_{i,n}+\D \sum_{\substack{
       h=1}}^s x_{h,n+1|r_h}+\sum_{\substack{
       h=s+1\\h\not=i}}^p x_{h,n+1|R_h}\right)+O(\varepsilon^2)\,. 
\end{equation}
}
\end{enumerate}

When using the induction hypothesis, we need to distinguish the case $ i\in\{1,\dots, s\}$ from the case $i\in \{s+1\,,\dots, p\}$. For the CML (\ref{equ1}), equations (\ref{eq12},\ref{eq121}) yield the following equations.
\begin{enumerate}
\item{For $i\in \{1,\dots,s\}$, we write the induction hypothesis for $m \ge 2$ as
\begin{align}\label{eq14}
	X_i(n+m) &= 
x_{i,n+m|r_i}+\left[ \D\frac{\partial f_{r_i}}{\partial r}(x_{i,n+m-1|r_i})+\D\sum_{k=n}^{n+m-2} \frac{\partial f_{r_i}}{\partial r}(x_{i,k|r_i})\prod_{l=k+1}^{n+m-1} \frac{\partial f_{r_i}}{\partial x}(x_{i,l|r_i})\right. \notag \\
	&\qquad + \left.\D\sum_{k=n+1}^{n+m-1} \left(\left(\sum_{\begin{subarray}{c}h=1\\h\not=i\end{subarray}}^s x_{h,k|r_h}+\sum_{\begin{subarray}{c}h=s+1\end{subarray}}^p x_{h,k|R_h}\right)\prod_{l=k}^{n+m-1} \frac{\partial f_{r_i}}{\partial x}(x_{i,l|r_i})\right)\right.\notag \\
	&\qquad\left. +\sum_{\begin{subarray}{c}h=1\\h\not=i\end{subarray}}^s x_{h,n+m|r_h}+\sum_{\begin{subarray}{c}h=s+1\end{subarray}}^p x_{h,n+m|R_h}\right]\varepsilon+O(\varepsilon^2)\,,
\end{align}
which implies that
\begin{align}
	X_i(n+m+1)
	&= f_{R_i}\left[ x_{i,n+m|r_i}+\left( \D\frac{\partial f_{r_i}}{\partial r}(x_{i,n+m-1|r_i})+\D\sum_{k=n}^{n+m-2} \frac{\partial f_{r_i}}{\partial r}(x_{i,k|r_i})\prod_{l=k+1}^{n+m-1} \frac{\partial f_{r_i}}{\partial x}(x_{i,l|r_i}) \right.\right.  \notag \\
	&\qquad + \left.\left.\D\sum_{k=n+1}^{n+m-1} \left(\left( \sum_{\begin{subarray}{c}h=1\\h\not=i\end{subarray}}^s x_{h,k|r_h}+\sum_{\begin{subarray}{c}h=s+1\end{subarray}}^p x_{h,k|R_h}\right)\prod_{l=k}^{n+m-1} \frac{\partial f_{r_i}}{\partial x}(x_{i,l|r_i})\right)
	\right.\right. \notag \\
	&\qquad + \left. \left.\sum_{\begin{subarray}{c}h=1\\h\not=i\end{subarray}}^s x_{h,n+m|r_h}+\sum_{\begin{subarray}{c}h=s+1\end{subarray}}^p x_{h,n+m|R_h}\right)\varepsilon\right] + \varepsilon \D \sum_{\begin{subarray}{c}h=1\\h\not=i \end{subarray}}^s f_{r_h}(x_{h,n+m|r_h}+O(\varepsilon))\notag \\
	&\qquad+ \varepsilon \D \sum_{\begin{subarray}{c}h=s+1\end{subarray}}^p f_{R_h}(x_{h,n+m|R_h}+O(\varepsilon))\notag \\\end{align}
	
We Taylor expand all occurrences of $f$ and its derivatives to obtain
\begin{align} \label{equ5bis}
	X_i(n+m+1)
	&=x_{i,n+m+1|r_i} + \left[\D\frac{\partial f_{r_i}}{\partial r}(x_{i,n+m|r_i})+
\D\sum_{k=n}^{n+m-1} \frac{\partial f_{r_i}}{\partial r}(x_{i,k|r_i})\prod_{l=k+1}^{n+m} \frac{\partial f_{r_i}}{\partial x}(x_{i,l|r_i}) \right. \notag \\
	&\qquad + \left.\D\sum_{k=n+1}^{n+m} \left(\left(\sum_{\begin{subarray}{c}h=1\\h\not=i\end{subarray}}^s x_{h,k|r_h}+\sum_{\begin{subarray}{c}h=s+1\end{subarray}}^p x_{h,k|R_h}\right)\prod_{l=k}^{n+m} \frac{\partial f_{r_i}}{\partial x}(x_{i,l|r_i})\right)\right. \notag \\
	&\qquad\left.+\D\sum_{\begin{subarray}{c}h=1\\h\not=i\end{subarray}}^s x_{h,n+m+1|r_h}+\D\sum_{\begin{subarray}{c}h=s+1\end{subarray}}^p x_{h,n+m+1|R_h}\right]\varepsilon +O(\varepsilon^2)\notag \,.
\end{align}
}

\item{
For $i\in \{s+1,\dots,p\}$, we write the induction hypothesis for $m \ge 2$ as
\begin{align}
\nonumber	X_i&(n+m)=x_{i,n+m|R_i}\\
	   &+\left[ \D\prod_{k=n}^{n+m-1} \frac{\partial f_{R_i}}{\partial x}(x_{i,k|R_i})A_{i,n}+\D\sum_{k=n+1}^{n+m-1} \left(\left(\D \sum_{\begin{subarray}{c}h=1 \end{subarray}}^s  x_{h,k|r_h}+ \sum_{\begin{subarray}{c}h=s+1\\h\not=i \end{subarray}}^p  x_{h,k|R_h}\right) \D\prod_{l=k}^{n+m-1} \frac{\partial f_{R_i}}{\partial x}(x_{i,l|R_i}) \right)\right. \notag \\
	&\qquad\left.   +\D\sum_{\begin{subarray}{c}h=1 \end{subarray}}^s x_{h,n+m|r_h}+\sum_{\begin{subarray}{c}h=s+1\\h\not=i \end{subarray}}^p x_{h,n+m|R_h}\right]\varepsilon+O(\varepsilon^2)\,,
\end{align}
which implies that
\begin{align}\label{equ51}
	X_i&(n+m+1) = \nonumber \\
	   &f_{R_i}\left( x_{i,n+m|R_i}+\left[ \D\prod_{k=n}^{n+m-1} \D\frac{\partial f_{R_i}}{\partial x} (x_{i,k|R_i})A_{i,n}+\D\sum_{k=n+1}^{n+m-1} \left(\left( \D \sum_{\begin{subarray}{c}h=1\end{subarray}}^s  x_{h,k|r_h} +\sum_{\begin{subarray}{c}h=s+1\\h\not=i \end{subarray}}^p  x_{h,k|R_h}\right) \right.\right. \right. \nonumber \\
	 &\qquad \left.\left.\left.\D\prod_{l=k}^{n+m-1} \D\frac{\partial f_{R_i}}{\partial x} (x_{i,l|R_i})\right)+ \D\sum_{\begin{subarray}{c}h=1 \end{subarray}}^s x_{h,n+m|r_h}+\sum_{\begin{subarray}{c}h=s+1\\h\not=i \end{subarray}}^p x_{h,n+m|R_h}\right]\varepsilon+O(\varepsilon^2)\right)  \notag \\
	 &\qquad +\varepsilon \D \sum_{\begin{subarray}{c}h=1 \end{subarray}}^s f_{r_h}(x_{h,n+m|r_h}+O(\varepsilon))+\sum_{\begin{subarray}{c}h=s+1\\h\not=i \end{subarray}}^p f_{R_h}(x_{h,n+m|R_h}+O(\varepsilon)) \notag \\
\end{align}

We Taylor expand of all occurrences of $f$ and its derivatives to obtain
\begin{align*}	
X_i(n+m+1)  &= x_{i,n+m+1|R_i}\\
	& +\left[ \D\prod_{k=n}^{n+m} \D\frac{\partial f_{R_i}}{\partial x} (x_{i,k|R_i})A_{i,n}+\D\sum_{k=n+1}^{n+m} \left( \D\left( \sum_{\begin{subarray}{c}h=1 \end{subarray}}^s  x_{h,k|r_h}+\sum_{\begin{subarray}{c}h=s+1\\h\not=i \end{subarray}}^p  x_{h,k|R_h}\right) \D\prod_{l=k}^{n+m} \D\frac{\partial f_{R_i}}{\partial x} (x_{i,l|R_i})  \right)\right.\\
	&\left.+\D\sum_{\begin{subarray}{c}h=1 \end{subarray}}^s x_{h,n+m+1|r_h}+\sum_{\begin{subarray}{c}h=s+1\\h\not=i \end{subarray}}^p x_{h,n+m+1|R_h}\right]\varepsilon+O(\varepsilon^2)
\end{align*}
}
\end{enumerate}
    
\qed

%%%%%%%%%%%%%%%%%%%%%%%%%%%%%%%%%%%%%%
%
%THEOREM 1
%%%%%%%%%%%%%%%%%%%%%%%%%

\subsection*{Theorem 1}

Let $|\varepsilon| < 1$ in the CML (\ref{equ1}), and suppose that the hypotheses of Lemma 1 are satisfied. That is, we assume that the map $f_{R_i}^{q_i}$ has an SN bifurcation at $R_i = r_i$, such that the associated SN orbit of $f_{r_i}$ is a {small-derivative} SN orbit, that $R_i=r_i + \varepsilon$ for $i\in\{1,\dots,s\}$, and that $R_i$ for $i\in\{s+1,\dots,p\}$ are far away from $r_i$.
Let $\{x_{i,1|r_i},\,x_{i,2|r_i},\dots , x_{i, q{_i}|r_i}\}$ be a period-$q_i$ orbit for the uncoupled oscillator $X_i$ for $ i\in \{1,\dots,s\}$, and let $\{x_{i,1|R_i},\,x_{i,2|R_i},\dots x_{i, q{_i}|R_i}\}$ be a period-$q_i$ orbit for the uncoupled oscillator $X_i$ for $ i\in \{s+1,\dots,p\}$. Consider the following initial conditions:
\begin{itemize}
\item For $i\in \{1,\dots,s\}$, let
\[		X_i(n) = x_{i,n|r_i}+\varepsilon A_{i,n} + O(\varepsilon^2)\,,\qquad
\]
where $x_{i,n|r_i}$ is the point of the SN orbit closest to the critical point of $f_{R_i}$ at $R_i=r_i$, and $A_{i,n}$ is an arbitrary $O(1)$ value.
\item For $i\in \{s+1,\dots,p\}$, let
\[		X_i(n) = x_{i,n|R_i}+\varepsilon A_{i,n} + O(\varepsilon^2)\,,\qquad
\]
where $x_{i,n|R_i}$ is a point of the orbit, and
\begin{align}
 	A_{i,n}&=\left[\D\D\sum_{k=n+1}^{n+q-1}\left(\left(\D\sum_{\substack{
       h=1}}^s  x_{h,k|r_h}+\sum_{\substack{
       h=s+1\\h\not=i}}^p  x_{h,k|R_h}\right)\D\prod_{l=k}^{n+q-1} \frac{\partial f_{R_i}}{\partial x}(x_{i,l|R_i})\right)\right.\nonumber\\
       &\left.+\D\sum_{\substack{
       h=1}}^s x_{h,n+q|r_h}+\D\sum_{\substack{
       h=s+1\\h\not=i}}^p x_{h,n+q|R_h}\right]
       \left(\frac{1}{1-\D\prod_{k=n}^{n+q-1}\frac{ \partial f_{R_i}}{\partial x}(x_{i,k|R_i})}\right)\,.
\end{align}
\end{itemize}

The CML (\ref{equ1}) has the solution
\begin{align}
		X_i(n+m) = 
			\begin{cases}
		x_{i,n+m|r_i}+\varepsilon A_{i,n+m}+O(\varepsilon^2)\,,\qquad
			i\in \{1,\dots,s\}\,,  m \in \{1, \dots,q\} \\
		x_{i,n+m|R_i}+\varepsilon A_{i,n+m}+O(\varepsilon^2)\,,\qquad
			i\in \{s+1,\dots,p\}\,,  m \in \{1, \dots,q\}\,,
			\end{cases}
\end{align}
where the coefficients $A_{i,n+m}$ are periodic with period $q=\mbox{lcm}(q_1,q_2,\dots,q_p)$ and satisfy the following formulas:
\begin{enumerate}
\item{For $ i\in \{1,\dots,s\}$,
\begin{align}\label{eqN5one}
A_{i,n+1} &= \frac{\partial f_{r_i}}{\partial r}(x_{i,n|r_i}) +\D \sum_{\substack{
      h=1\\h\not=i}}^s x_{h,n+1|r_h}+\sum_{\substack{
       h=s+1}}^p x_{h,n+1|R_h}\,, \quad m = 1 \\
\end{align}
%and for $m \geq 2$,
\begin{align}\label{eqN5}
	A_{i,n+m} &= \D\frac{\partial f_{r_i}}{\partial r}(x_{i,n+m-1|r_i})+\D\sum_{k=n}^{n+m-2} \frac{\partial f_{r_i}}{\partial r}(x_{i,k|r_i})\prod_{l=k+1}^{n+m-1}  \frac{\partial f_{r_i}}{\partial x}(x_{i,l|r_i}) \notag \\
		&\qquad +\D\sum_{k=n+1}^{n+m-1} \left(\D\left(\sum_{\begin{subarray}{c}h=1\\h\not=i \end{subarray}}^s x_{h,k|r_h}+\sum_{\begin{subarray}{c}h=s+1\end{subarray}}^p x_{h,k|R_h}\right)\prod_{l=k}^{n+m-1} \frac{\partial f_{r_i} }{\partial x}(x_{i,l|r_i})\right) \notag \\
		&\qquad +\sum_{\begin{subarray}{c}h=1\\h\not=i \end{subarray}}^s x_{h,n+m|r_h}+\sum_{\begin{subarray}{c}h=s+1\end{subarray}}^p x_{h,n+m|R_h}\,, \qquad m \in \{2, \dots,q\}\,.
\end{align}
}
\item{For  $i\in\{s+1,\dots,p\}$, 
\begin{align}\label{eqN6}
 	A_{i,n+m}&=\left[\D\D\sum_{k=n+m+1}^{n+m+q-1}\left(\left(\D\sum_{\substack{
       h=1}}^s  x_{h,k|r_h}+\sum_{\substack{
       h=s+1\\h\not=i}}^p  x_{h,k|R_h}\right)\D\prod_{l=k}^{n+m+q-1} \frac{\partial f_{R_i}}{\partial x}(x_{i,l|R_i})\right)\right.\nonumber\\
       &\left.+\D\sum_{\substack{
       h=1}}^s x_{h,n+m+q|r_h}+\D\sum_{\substack{
       h=s+1\\h\not=i}}^p x_{h,n+m+q|R_h}\right]
       \left(\frac{1}{1-\D\prod_{k=n+m}^{n+m+q-1}\frac{ \partial f_{R_i}}{\partial x}(x_{i,k|R_i})}\right)\,,\nonumber\\
       & \qquad m \in \{1, \dots,q\}\,.
\end{align}
}
\end{enumerate}

\paragraph*{Remark}
Although the initial conditions given in the statement of Theorem 1 may seem restrictive, our numerical computations demonstrate that --- independently of the type of the orbit (i.e., either close to or far away from the SN) --- it is sufficient to take as an initial condition any point of the unperturbed orbit plus a perturbation of size $O(\varepsilon)$.

%%%%%%%%%%%%%

\subsection*{Proof of Theorem 1.}

We need to consider $i \in \{ 1, \dots, s\}$ and $i \in \{ s+1, \dots, p\}$ separately.

\begin{enumerate}
\item

Using Lemma 1, it follows from $X_i(n)=x_{i,n|r_i}+\varepsilon A_{i,n}+O(\varepsilon^2)$ that
     \begin{align}\label{equM1}
X_i(n+1)=x_{i,n+1|r_i}+\varepsilon \left(\D \frac{\partial f_{r_i}}{\partial r}(x_{i,n|r_i}) +\D \sum_{\substack{
       h=1\\h\not=i}}^s x_{h,n+1|r_h} +\sum_{\substack{
       h=s+1}}^p x_{h,n+1|R_h}\right)+O(\varepsilon^2) 
       \end{align}
           and
       \begin{align}\label{equN1}
        X_i(n+q+1)
          &=x_{i,n+q+1|r_i} + \left[\D\frac{\partial f_{r_i}}{\partial r}(x_{i,n+q|r_i})+
\D\sum_{k=n}^{n+q-1} \frac{\partial f_{r_i}}{\partial r}(x_{i,k|r_i})\prod_{l=k+1}^{n+q} \frac{\partial f_{r_i}}{\partial x}(x_{i,l|r_i}) \right. \notag \\
	&\qquad + \left.\D\sum_{k=n+1}^{n+q} \left(\left(\sum_{\begin{subarray}{c}h=1\\h\not=i\end{subarray}}^s x_{h,k|r_h}+\sum_{\begin{subarray}{c}h=s+1\end{subarray}}^p x_{h,k|R_h}\right)\prod_{l=k}^{n+q} \frac{\partial f_{r_i}}{\partial x}(x_{i,l|r_i})\right)\right. \notag \\
	&\qquad+\left.\D\sum_{\begin{subarray}{c}h=1\\h\not=i\end{subarray}}^s x_{h,n+q+1|r_h}+\sum_{\begin{subarray}{c}h=s+1\end{subarray}}^p x_{h,n+q+1|R_h}\right]\varepsilon +O(\varepsilon^2)\,.
\end{align}

Because $\D\prod_{l=k}^{n+q} \frac{\partial f_{r_i}}{\partial x}(x_{i,l|r_i})$ includes the arbitrarily small term
$\left|\D\frac{\partial f_{r_i}(x_{i,n|r_i})}{\partial x}\right|$, it follows from (\ref{equN1}) that
\begin{equation*}
	X_i(n+q+1)=x_{i,n+q+1|r_i}+\left(\D\frac{\partial f_{r_i}}{\partial r}(x_{i,n+q|r_i})+\D\sum_{\begin{subarray}{c}h=1\\h\not=i\end{subarray}}^s x_{h,n+q+1|r_h}+\sum_{\begin{subarray}{c}h=s+1\end{subarray}}^p x_{h,n+q+1|R_h}\right)\varepsilon+O(\varepsilon^2)\,.
\end{equation*}

With $q=\mbox{lcm}(q_1,q_2,\dots,q_p)$, we have 
\begin{align*}
	x_{i,n+q+1|r_i} &= x_{i,n+1|r_i} \,, \\
	x_{i,n+q+1|R_i} &= x_{i,n+1|R_i}\,, \\
	\D\sum_{\begin{subarray}{c}h = 1\\h\not=i\end{subarray}}^s x_{h,n+q+1|r_h} +\sum_{\begin{subarray}{c}h=s+1\\h\not=i\end{subarray}}^p x_{h,n+q+1|R_h} &= \D \sum_{\substack{
       h=1\\h\not=i}}^s x_{h,n+1|r_h}+\sum_{\substack{
       h=s+1\\h\not=i}}^p x_{h,n+1|R_h}\,,
\end{align*}        
because $x_{i,j}$ is a point of a $q_i$-period orbit. Consequently, equations (\ref{equM1}) and (\ref{equN1}) become the same equation. From equations (\ref{eq12}) and (\ref{equM1}), we can write equation (\ref{one}) in Lemma 1 as $X_i(n+m)=x_{i,n+m|r_i}+\varepsilon A_{i,n+m}+O(\varepsilon^2)$ to obtain 
\begin{align}
	A_{i,n+m} &= \D\frac{\partial f_{r_i}}{\partial r}(x_{i,n+m-1|r_i})+\D\sum_{k=n}^{n+m-2} \frac{\partial f_{r_i}}{\partial r}(x_{i,k|r_i})\prod_{l=k+1}^{n+m-1}  \frac{\partial f_{r_i}}{\partial x}(x_{i,l|r_i}) \notag \\
		&\qquad +\D\sum_{k=n+1}^{n+m-1} \left(\D\left(\sum_{\begin{subarray}{c}h=1\\h\not=i \end{subarray}}^s x_{h,k|r_h}+\sum_{\begin{subarray}{c}h=s+1\end{subarray}}^p x_{h,k|R_h}\right)\prod_{l=k}^{n+m-1} \frac{\partial f_{r_i} }{\partial x}(x_{i,l|r_i})\right) \notag \\
		&+\sum_{\begin{subarray}{c}h=1\\h\not=i \end{subarray}}^s x_{h,n+m|r_h}+\sum_{\begin{subarray}{c}h=s+1\end{subarray}}^p x_{h,n+m|R_h}\,,\qquad m \in \{1, \dots,q\}\,.
\end{align}

\item
With $X_i(n+m)=x_{i,n+m|R_i}+\varepsilon A_{i,n+m}+O(\varepsilon^2)$, Lemma 1 implies that
\end{enumerate}
    \begin{align}
     X_i(n+m+q) &=x_{i,n+m+q|R_i}+\left[ \D\prod_{k=n+m}^{n+m+q-1} \frac{\partial f_{R_i}}{\partial x}(x_{i,k|R_i})A_{i,n+m}\right. \nonumber \\
                            &\quad +\left.\D\sum_{k=n+m+1}^{n+m+q-1} \left(\left( \D\sum_{\substack{h=1}}^s x_{h,k|r_h}+ \sum_{\substack{h=s+1\\h\not=i}}^p  x_{h,k|R_h}\right)\D\prod_{l=k}^{n+m+q-1} \frac{\partial f_{R_i}}{\partial x}(x_{i,l|R_i})\right)   \right.  \nonumber \\  
                            &\left.+\D\sum_{\substack{h=1}}^s x_{h,n+m+q|r_h}+\sum_{\substack{h=s+1\\h\not=i}}^p x_{h,n+m+q|R_h}\right]\varepsilon+O(\varepsilon^2)\,. \nonumber 
    \end{align}
    
          By taking $q=\mbox{lcm}(q_1,q_2,\dots,q_p)$, we obtain $x_{i,n+m|r_i}=x_{i,n+m+q|r_i}$ and  $x_{i,n+m|R_i}=x_{i,n+m+q|R_i}$  because $x_{i,j}$ is a point of a periodic orbit. Consequently, $X_i(n+m)-X_i(n+m+q)=O(\varepsilon)$ whenever      
        \begin{align}\label{eqN4}    
          A_{i,n+m}&= \D\prod_{k=n+m}^{n+m+q-1} \frac{\partial f_{R_i}}{\partial x}(x_{i,k|R_i})A_{i,n+m}+\D\sum_{k=n+m+1}^{n+m+q-1} \left( \D \left(\sum_{\substack{
       h=1}}^s  x_{h,k|r_h} +\sum_{\substack{h=s+1\\h\not=i}}^p  x_{h,k|R_h}\right) \right.\nonumber\\
       &\left.\D\prod_{l=k}^{n+m+q-1} \frac{\partial f_{R_i}}{\partial x}(x_{i,l|R_i})\right)+\D\sum_{\substack{
       h=1}}^s x_{h,n+m+q|r_h}+\sum_{\substack{
       h=s+1\\h\not=i}}^p x_{h,n+m+q|R_h}\,.
       \end{align}
Furthermore, $A_{i,n+m}$ is periodic.  

Equation (\ref{eqN4}) now implies that   
 \begin{align}
 	A_{i,n+m}&=\left[\D\D\sum_{k=n+m+1}^{n+m+q-1}\left(\left(\D\sum_{\substack{
       h=1}}^s  x_{h,k|r_h}+\sum_{\substack{h=s+1\\h\not=i}}^p  x_{h,k|R_h}\right)
       \D\prod_{l=k}^{n+m+q-1} \frac{\partial f_{R_i}}{\partial x}(x_{i,l|R_i})\right)\right.\nonumber\\
       &\left.+\D\sum_{\substack{
       h=1}}^s x_{h,n+m+q|r_h}+\D\sum_{\substack{
       h=s+1\\h\not=i}}^p x_{h,n+m+q|R_h}\right]
       \left(\frac{1}{1-\D\prod_{k=n+m}^{n+m+q-1}\frac{ \partial f_{R_i}}{\partial x}(x_{i,k|R_i})}\right)\,,\nonumber\\
       & \qquad m \in \{1, \dots,q\}\,.
\end{align}       
It follows that $A_{i,n+m}$ has period $q$ because it is given by sums and products of $q$-periodic functions evaluated at %sequential 
points of a $q$-periodic orbit.

\qed

Observe that the formula for $A_{i,j}$ for $i\in \{1,\dots,s\}$ in equation (\ref{eqN5}) does not contain the term
$\left[1-\D\prod_{k=j}^{j+q-1} \frac{\partial f_{r_i}}{\partial x}(x_{i,k|r_i})\right]$ in the denominator [see equation (\ref{12A})]. Otherwise, $A_{i,j}$ would be of size $O({1}/{\varepsilon})$, and the expansion that we used to prove Theorem 1 would not be valid. By contrast, the formula for $A_{i,j}$ for $i \in \{s+1,\dots,p\}$ in equation (\ref{eqN6}) includes the term $\left[1-\D\prod_{k=j}^{j+q-1} \frac{\partial f_{R_i}}{\partial x}(x_{i,k|R_i})\right]$ in the denominator because the oscillators are far from SN bifurcations for $i\in\{s+1,\dots,p\}$. Therefore, 
\begin{equation*}
	1-\D\prod_{k=j}^{j+q-1} \frac{\partial f_{R_i}}{\partial x}(x_{i,k|R_i}) = O(1)\,, 
\end{equation*}	
and it follows that $A_{i,j}$ also has size $O(1)$.

%%%%%

\subsection{Type-I Intermittency Near Saddle--Node Bifurcations}\label{sub:TypeI}

Theorem 1 concerns the behavior of the CML (\ref{equ1}) with a mixture of periodic oscillators that are near the SN bifurcation with others that are far from the SN bifurcation. If an SN orbit takes place at $R_i=r_i$, then the oscillators with $R_i=r_i+\varepsilon$ are the ones that are close to the SN orbit. 

We now want to study the behavior of the CML (\ref{equ1}) when an uncoupled oscillator has type-I intermittency \cite{pomeau} at $R_i=r_i-\varepsilon$ (i.e., just to the left of where it undergoes an SN bifurcation). Type-I intermittency is characterized by the alternation of an apparently periodic regime (a so-called ``laminar phase''), whose mean duration follows the power law $\langle l \rangle \propto \varepsilon^{-\frac{1}{2}}$ (so the laminar region becomes longer as $\varepsilon$ becomes smaller), and chaotic bursts. As $R_i=r_i-\varepsilon$, we expand $f_{r_i-\varepsilon}$ in powers of $\varepsilon$ to obtain
\begin{equation*}
	f_{r_i-\varepsilon}^{q_i}(x_{j|r_i})=f_{r_i}^{q_i}(x_{j|r_i})-\varepsilon  \frac{\partial f_{r_i}^{q_i}}{\partial r}(x_{j|r_i}) + O(\varepsilon^2)\,,
\end{equation*}	
where $x_j$ a point of a period-$q_i$ SN orbit. Therefore the laminar phase is driven by the period-$q_i$ SN orbit associated with the SN bifurcation. Thus, as $\varepsilon$ becomes smaller, the orbit spends more iterations in the laminar regime,
 %(i.e., with $\langle l \rangle \propto \varepsilon^{-\frac{1}{2}}$) 
 and it thus more closely resembles the period-$q_i$ SN orbit. In particular, $\left|x_{j|r_i}-f_{r_{i-\varepsilon}}^{q_i}(x_{j|r_i})\right| = O(\varepsilon)\,$. 
 
To approximate the temporal evolution of the laminar regime using the  period-$q_i$ SN orbit, we proceed in the same way as in Theorem 1, except that we replace $R_i=r_i+\varepsilon$ by $R_i=r_i-\varepsilon$.  We thus write
\begin{align}
X_i(n+1) &= x_{i,n+1|r_i} + \left[ -\D\frac{\partial f_{r_i}}{\partial r}(x_{i,n|r_i}) + \sum_{\substack{
            h=1\\h\not=i}}^s x_{h,n+1|r_h}+\sum_{\substack{
            h=s+1}}^p x_{h,n+1|R_h}\right]\varepsilon+O(\varepsilon^2)\,, \quad m = 1\,,
\end{align}
%and for $m \geq 2$
\begin{align}
X_i(n+m) &= x_{i,n+m|r_i} + \left[ -\D\frac{\partial f_{r_i}}{\partial r}(x_{i,n+m-1|r_i})-\D\sum_{k=n}^{n+m-2} \frac{\partial f_{r_i}}{\partial r}(x_{i,k|r_i})\prod_{l=k+1}^{n+m-1} \frac{\partial f_{r_i}}{\partial x}(x_{i,l|r_i})\right. \notag \\
&\qquad \left.+\D\sum_{k=n+1}^{n+m-1} \left(\left(\sum_{\substack{h=1\\h\not=i}}^s x_{h,k|r_h}+\sum_{\substack{h=s+1}}^p x_{h,k|R_h}\right)\prod_{l=k}^{n+m-1} \frac{\partial f_{r_i}}{\partial x}(x_{i,l|r_i})\right)\right.\\
            &\qquad \left.+\sum_{\substack{
            h=1\\h\not=i}}^s x_{h,n+m|r_h}+\sum_{\substack{
            h=s+1}}^p x_{h,n+m|R_h}\right]\varepsilon+O(\varepsilon^2)\,, \quad m \in \{2,\dots, q\}\,
\end{align}
 which determines the temporal evolution of the CML (\ref{equ1}) in the laminar regime.

%%%%%

\section{Numerical Computations}\label{sec:numerical}

Theorem 1 proves the existence of an approximately periodic orbit. In principle, one can deduce the existence of a periodic orbit by using the Implicit Function Theorem (IFT). However, the IFT fails at the SN bifurcation (i.e., at $R_i=r_i$) for free oscillators and consequently fails near an SN bifurcation (i.e., for $R_i=r_i+\varepsilon$) of the HWCML (\ref{equ1}), because the 
%coupling term $\varepsilon$ causes the 
Jacobian determinant vanishes.

Had we expanded all terms in Theorem 1, we would have obtained terms of size $O(\varepsilon^2)$ that depend on the coefficients of the terms of size $O(\varepsilon)$ (i.e., as functions of the $A_{i,n+m}$ terms in Theorem 1), so terms of size $O(\varepsilon^2)$ would have the same period as the $A_{i,n+m}$. We could then obtain terms of size $O(\varepsilon^3)$ as a functions of the coefficients of lower-order terms. These terms would also have the same period as $A_{i,n+m}$, and the same is true for all higher-order terms if we continued the expanding in powers of $\varepsilon$. This reasoning suggests the existence of a periodic orbit of period $q=\mbox{lcm}(q_1, \dots ,q_p)$ (not just an approximate one), and our numerical simulations successfully illustrate the existence of such periodic orbits.

For simplicity, we consider a pair of coupled oscillators,
\begin{align}\label{equA}
	X(n+1) &= f(X(n))+\varepsilon g(Y(n)) \,, \notag \\
	Y(n+1) &= g(Y(n))+\varepsilon f(X(n))\,,
\end{align}
where $f(x)=R_{1}x(1-x)$ and $g(y)=\cos(R_{2}y)$. We initially fix the coupling to be $\varepsilon=0.0001$, though we will later consider $2\varepsilon$, $3\varepsilon$, and so on. The uncoupled oscillator $Y(n)$ has a fixed period of $4$ and is far away from a SN bifurcation for $R_{2}=1.9$. We use values of $R_1$ such that the uncoupled oscillator $X(n)$ is near an SN bifurcation, and we consider SN orbits with different periods.

%%%%%

\subsection{Uncoupled Oscillator $X(n)$ with a Period-3 Orbit} \label{per3}

\begin{figure}[!hhh]
    \includegraphics[width=\textwidth]{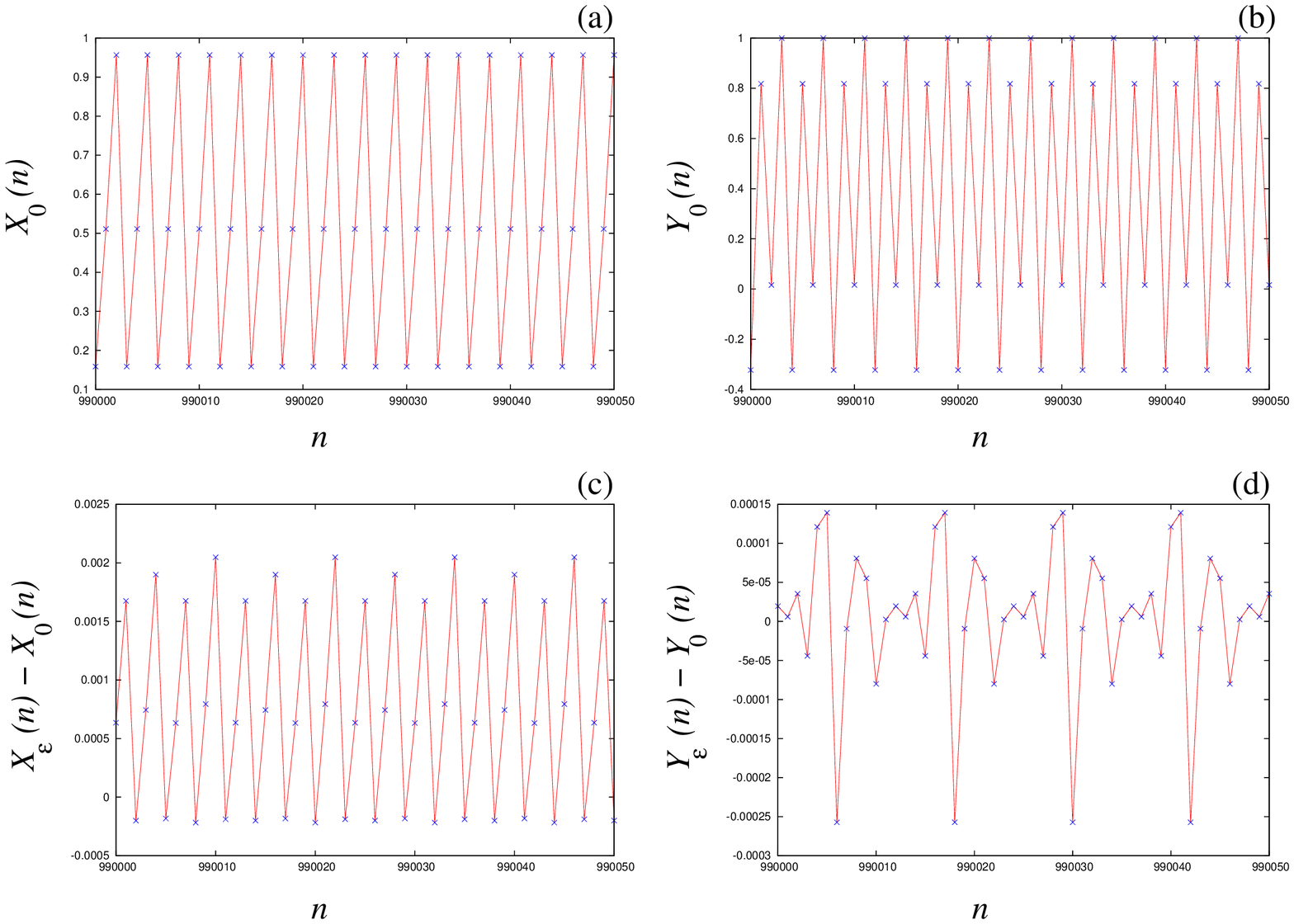}
	\caption{\label{fig:SNab}
             {\bf Temporal evolution of the HWCML (\ref{equA}) for $R_1 = r_1 + 2\varepsilon$, where $r_1 \approx 3.828427$ (which is an SN bifurcation point) and $R_2 = 1.9$}.
             The uncoupled oscillators have (a) period 3 and (b) period 4. When $\varepsilon=0.0001$ (i.e., weak coupling), the oscillators $X_\varepsilon(n)$ and $Y_\varepsilon(n)$ both have period $\lcm(3,4)=12$.  In panels (c) and (d), we plot $X_\varepsilon(n) - X_0(n)$ and $Y_\varepsilon(n) - Y_0(n)$ (i.e., the solution in the coupled case minus the solution in the $\varepsilon = 0$ case) to better observe the period-12 dynamics.
	}
\end{figure}

For the oscillator $X(n)$, we fix $R_1=r_1+2\varepsilon$, where $r_1 \approx 3.828427$ is an SN bifurcation point of $f$. When there is no coupling, the free oscillator $X(n)$ has a period-3 SN orbit, and the free oscillator $Y(n)$ has a period-$4$ orbit. When coupled, both $X(n)$ and $Y(n)$ have a periodic orbit with period $q=\lcm(3,4)=12$ (see Fig.~\ref{fig:SNab}).
%, as we have previously predicted at the beginning of the section. This is shown in Fig.~\ref{fig:SNab}.

\begin{figure}[!hhh]
    \includegraphics[width=\textwidth]{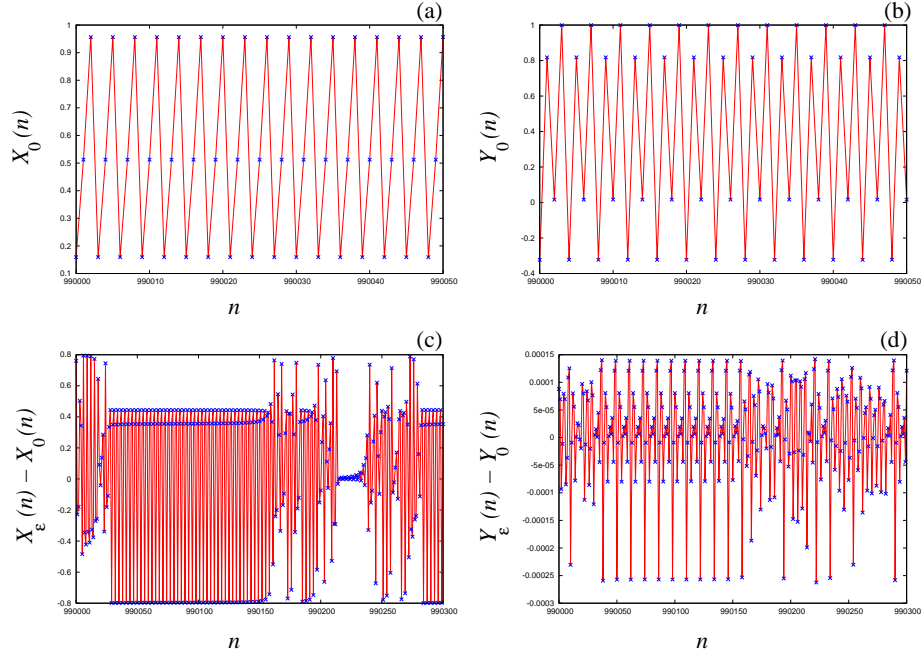}
	\caption{\label{fig:SN2ab}
             {\bf Temporal evolution of the HWCML (\ref{equA}) for $R_1 = r_1 + \varepsilon$, where $r_1 \approx 3.828427$ (which is an SN bifurcation point) and $R_2 = 1.9$.}
The uncoupled oscillators have (a) period 3 and (b) period 4. When $\varepsilon=0.0001$ (i.e., weak coupling), the oscillators $X_\varepsilon(n)$ and $Y_\varepsilon(n)$ exhibit type-I intermittency. In panels (c) and (d), we plot $X_\varepsilon(n) - X_0(n)$ and $Y_\varepsilon(n) - Y_0(n)$ (i.e., the solution in the coupled case minus the solution in the $\varepsilon = 0$ case) to better observe the intermittency dynamics.
	}
\end{figure}

At $R_1=r_1+\varepsilon$, the HWCML (\ref{equA}) exhibits type-I intermittency associated with the SN bifurcation (see Fig.~\ref{fig:SN2ab}). However, for larger $R_1$ (e.g., $r_1+2\varepsilon$, $r_1+3\varepsilon$, $\dots$, $r_1+7\varepsilon$),
%, and so %on, t
the periods of the uncoupled oscillators $X(n)$ and $Y(n)$ are preserved
%(as expected)
because we are farther away from the bifurcation point. We observe type-I intermittency for $R_1 = r_1+0\varepsilon$, $R_1 = r_1-\varepsilon$, $R_1 = r_1-2\varepsilon$.

\medskip

\paragraph*{Remark} 
When $R_1 = r_1 + 2 \varepsilon$, we calculate $1-\prod_{k=j}^{j+m-1}\frac{\partial f_{r_i}}{\partial x}(x_{i,k|r_i}) \approx 0.24$ for $\varepsilon = 0.0001$. (For $R_1 = r_1 + \varepsilon$, we obtain a smaller value for the second quantity). Recall the quantifications of ``far from'' and ``near'' in section \ref{sub:PeriodicOrbits}. Although $\varepsilon$ can be very small, the periodic windows that are born with an SN orbit can be even smaller than $\varepsilon$. Thus, from the dynamical standpoint, a very small value of the coupling parameter $\varepsilon$ can nevertheless be large as a variation on a bifurcation parameter. 

\medskip

\begin{figure}[!hhh]
\begin{centering}
    \includegraphics[width=0.5\textwidth]{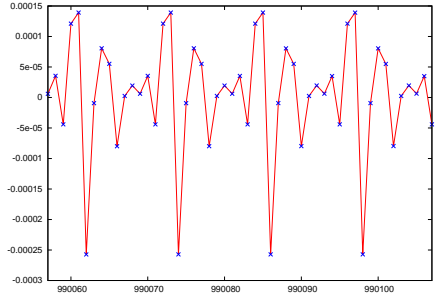}
	\caption{\label{fig:SN2amp}
            {\bf Magnification of the laminar regime of type-I intermittency from Fig.~\ref{fig:SN2ab}d}. We can clearly see the resemblance with the temporal evolution of the oscillator in the periodic regime (see Fig.~\ref{fig:SNab}d).
	}
	\end{centering}
\end{figure}

In Section \ref{sub:TypeI}, we determined the temporal evolution of the oscillators in the laminar regime of type-I intermittency up to size $O(\varepsilon)$. By comparing Fig.~\ref{fig:SNab} (which depicts the dynamics for a parameter value slightly larger than the SN bifurcation point) and Fig.~\ref{fig:SN2amp} (which depicts the dynamics right before the bifurcation), we observe periodic behavior just after the bifurcation and laminar behavior just before it.

%%%%

\subsection{Uncoupled Oscillator $X(n)$ with a Period-5 Orbit} \label{per5}

We proceed as in Section \ref{per3} and obtain similar results.

\begin{figure}[!hhh]
    \includegraphics[width=\textwidth]{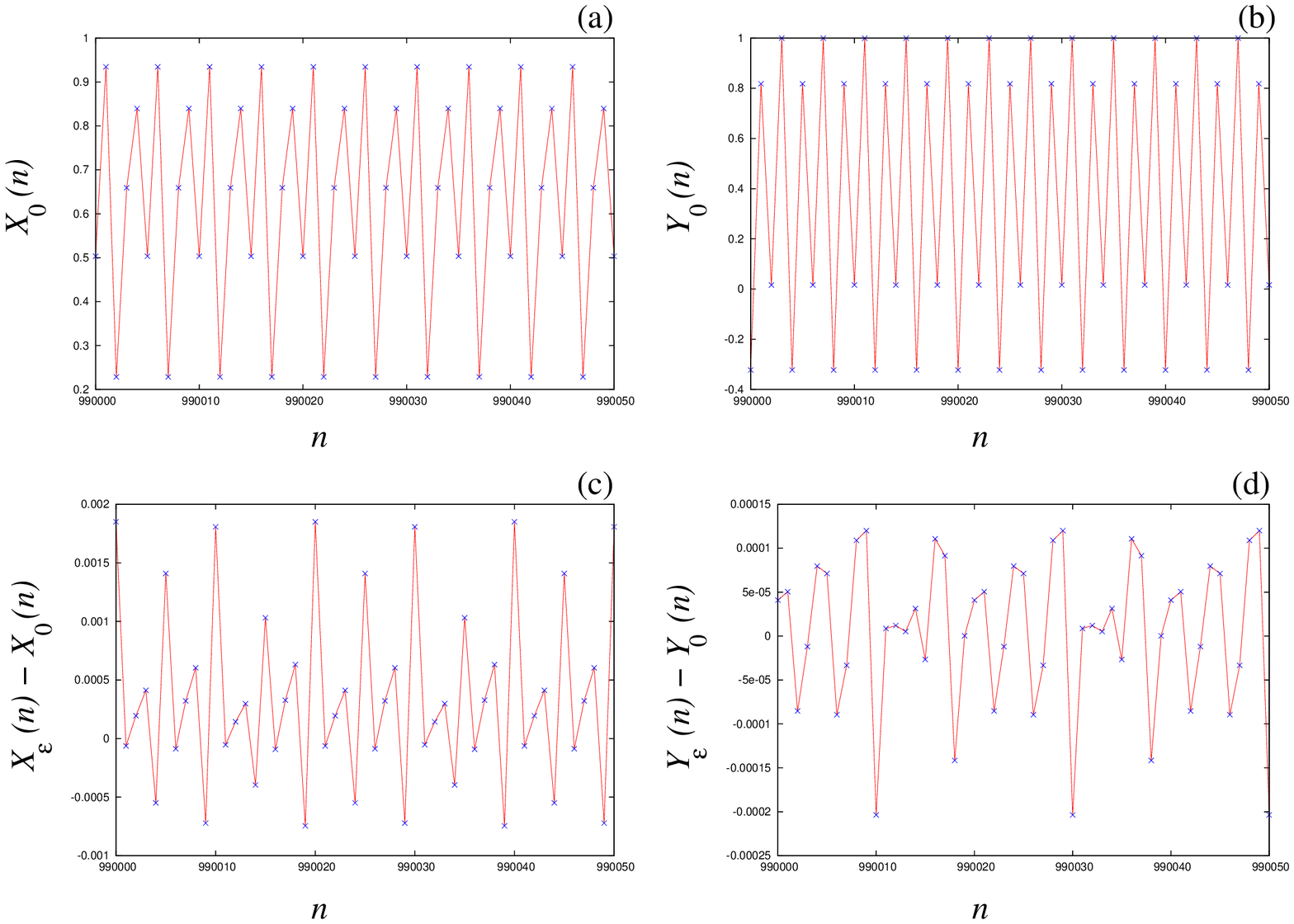}
	\caption{\label{fig:SNcd}
             {\bf Temporal evolution of the HWCML (\ref{equA}) for $R_1 = r_1 + 2\varepsilon$, where $r_1\approx3.738173$ (which is an SN bifurcation point) and $R_2 = 1.9$.}
The uncoupled oscillators have (a) period 5 and (b) period 4. When $\varepsilon=0.0001$ (i.e., weak coupling), the oscillators $X_\varepsilon(n)$ and $Y_\varepsilon(n)$ both have period $\lcm(5,4)=20$.  In panels (c) and (d), we plot $X_\varepsilon(n) - X_0(n)$ and $Y_\varepsilon(n) - Y_0(n)$ (i.e., the solution in the coupled case minus the solution in the $\varepsilon = 0$ case) to better observe the period-20 dynamics.
	}
\end{figure}

For the oscillator $X(n)$, we fix $R_1=r_1+2\varepsilon$, where $r_1 \approx 3.738173$ is an SN bifurcation point of $f$. When there is no coupling, the free oscillator $X(n)$ has a period-5 SN orbit, and the free oscillator $Y(n)$ has a period-$4$ orbit. When coupled, both $X(n)$ and $Y(n)$ have a periodic orbit with period $q=\lcm(5,4)=20$ (see Fig.~\ref{fig:SNcd}).

\begin{figure}[!hhh]
    \includegraphics[width=\textwidth]{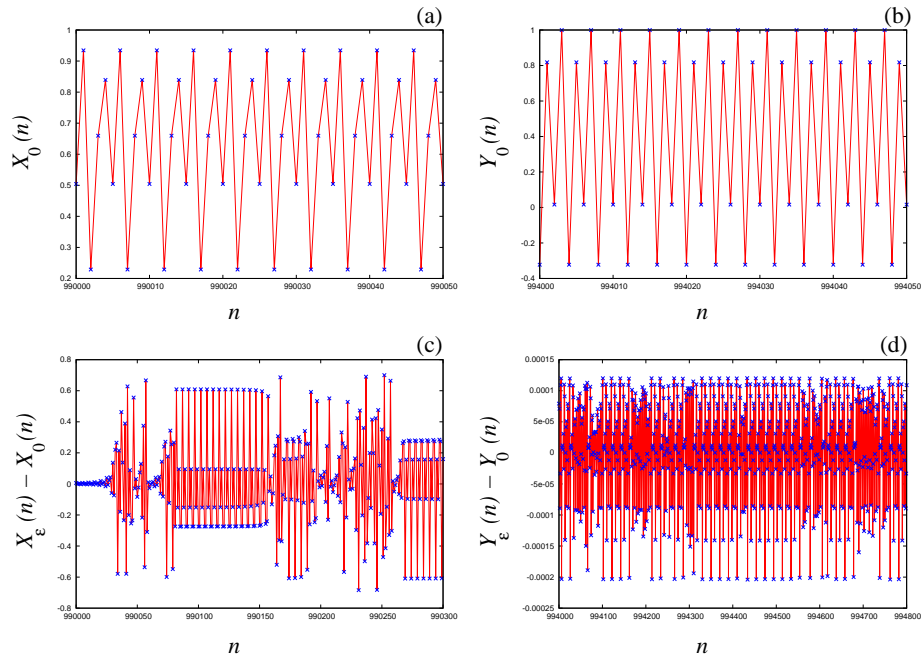}
	\caption{\label{fig:SN2cd}
             {\bf Temporal evolution of the HWCML (\ref{equA}) for $R_1 = r_1 + \varepsilon$, where $r_1$ is an SN bifurcation point and $R_2 = 1.9$}.
             The uncoupled oscillators have (a) period 5 and (b) period 4. When $\varepsilon=0.0001$ (i.e., weak coupling), the oscillators $X_\varepsilon(n)$ and $Y_\varepsilon(n)$ show intermittency. In panels (c) and (d), we plot $X_\varepsilon(n) - X_0(n)$ and $Y_\varepsilon(n) - Y_0(n)$ (i.e., the solution in the coupled case minus the solution in the $\varepsilon = 0$ case) to better observe the intermittent dynamics.
	}
\end{figure}

At $R_1=r_1+\varepsilon$, the HWCML (\ref{equA}) exhibits type-I intermittency associated with the SN bifurcation (see Fig.~\ref{fig:SN2cd}). However, for larger $R_1$ (e.g., $r_1+2\varepsilon$, $r_1+3\varepsilon$, $r_1+4\varepsilon$ \dots), the periods of the uncoupled oscillators $X(n)$ and $Y(n)$ are preserved because we are farther away from the bifurcation point. We observe type-I intermittency for $R_1 = r_1+0\varepsilon$, $R_1 = r_1-\varepsilon$, $R_1 = r_1-2\varepsilon$.

%%%%%

\subsection{Summary of HWCML Dynamics}\label{alp}
% Illustrated in Numerical Simulations

Our results allow us to deduce the dynamics of the HWCML (\ref{equA}) when $R_i=r_i+\varepsilon^\alpha$. We worked with a coupling strength of $\varepsilon=0.0001$ and a control parameter of $R_i=r_i+k\varepsilon$. In our numerical computations, we observed the following behavior:
\begin{enumerate}
	\item[(a)] intermittency for $R_i \leq r_i + \varepsilon$;
	\item[(b)] periodic behavior for $R_i \geq r_i + 2 \varepsilon$.
\end{enumerate}
Therefore, the following occurs.
\begin{enumerate}
	\item[(i)] If we choose $R_i=r_i + \varepsilon^\alpha$ with $\alpha > 1$, then $R_i < r_i + \varepsilon$, and the HWCML exhibits intermittent behavior according to (a).
	\item[(ii)] If we choose $R_i=r_i + \varepsilon^\alpha$ with $0 < \alpha < 1$, then $R_i > r_i + 2\varepsilon$; this holds even for $\alpha$ close to $1$, as long as $\varepsilon^\alpha > 2\varepsilon$ (e.g., $0 <\alpha \lessapprox 0.92$ for $\varepsilon = 0.0001$). Therefore, the HWCML exhibits periodic behavior according to (b).
\end{enumerate}

Based on our numerical computations, we can thus establish the following statement: ``Under the hypotheses of Theorem 1, the oscillators in the CML (\ref{equ1}) have periodic orbits that persist with the same period as in Theorem 1 for perturbations of size $O(\varepsilon)$. That is, higher-order terms do not change the period, as we heuristically stated at the beginning of Section \ref{sec:numerical}.''

We now discuss the consequences of all oscillators in an HWCML having the same period $q=\mbox{lcm}(q_1,q_2,\dots,q_p)$, where $q_1,\dots, q_p$ are the periods of the free oscillators. One can adjust the parameters to obtain periods ${q_1\dots q_p}$ so that $q = \lcm(q_1, q_2 , \dots , q_p)$ remains constant. For example, if $q_1=3$ and $q_2=2^k$, then $q=\lcm(3,2^k)=3 \times 2^k$ (for integers $k >0$). If the first oscillator undergoes a period-doubling cascade, then its period is $3$, $3 \times 2$, $3 \times 2^2$, and so on. However, the period $m$ of the HWCMLs is $q = \lcm(3,2^k)= \lcm(3 \times 2,2^k)= \dots = \lcm(3 \times 2^k,2^k)=3 \times 2^k$, so it does not change even after an arbitrary number of period-doubling bifurcations.  That is, for arbitrarily small $\varepsilon \neq 0$, the HWCML period remains the same even amidst a period-doubling cascade. 

We illustrate the above phenomenon with a simple computation. Consider the HWCML (\ref{equA}) and suppose that $R_{1}=3.83$ and $R_{2}=1.9$.  When $\varepsilon=0$ (i.e., when there is no coupling), the free oscillator $X(n)$ has a period-3 orbit and the free oscillator $Y(n)$ has a period-$4$ orbit. However, when $\varepsilon=0.001$, both $X(n)$ and $Y(n)$ have a periodic orbit with period $q=\lcm(3,4)=12$. As we show in Table \ref{tab:TABLA1}, the free oscillator $X(n)$ undergoes period-doubling bifurcations, but the HWCML exhibits synchronization and still has period-12 orbits for $\varepsilon=0.001$.

%%%%%

\section{Conclusions and Discussion}\label{discuss}

We have examined heterogeneous weakly coupled map lattices (HWCMLs) and have given results to describe periodic orbits both near and far from saddle--node orbits and to describe the temporal evolution of the laminar regime in type-I intermittency. All periodic windows of the bifurcation diagram of unimodal maps originate from SN bifurcations, so it is important to explore the dynamics near such bifurcation points.

An important implication of our results is that HWCMLs of oscillators need not behave approximately like their associated free-oscillator counterparts. In particular, they can have periodic-orbit solutions with completely different periods even for arbitrarily small coupling strengths $\varepsilon \neq 0$.  

Our numerical calculations illustrate an important result about period preservation when oscillator parameters change. Even when one varies the parameters $R_i$ of the functions $f_{R_i}$ such that the uncoupled oscillator $X_i$ undergoes a period-doubling cascade, the periods of each of the coupled oscillators are preserved as long as the least common multiple of the periods remains constant. That is, the oscillation period is resilient to changes.

Period preservation is a rather generic phenomenon in CMLs. Suppose, for example, that one oscillator has period of $q \times 2^n$, which can originate either from period doubling or from an SN bifurcation \cite{SanMartin2007}. One can then change parameters so that different individual oscillators (if uncoupled) would undergo a period-doubling cascade, whereas the least common multiple of the periods of those oscillators will remain constant until one oscillator (if uncoupled) has period $p \times 2^{n+1}$. In a CML, a very large number of oscillators can each undergo a period-doubling cascade, so the period of a CML can be very resilient even in situations when other conditions --- in particular, the values of the parameters in the CML --- are changing a lot. Moreover, one can adjust the parameters to obtain oscillations of arbitrary periods ${q_1\dots q_p}$ with $q = \lcm(q_1, q_2 , \dots , q_p) = \mathrm{constant}$. Consequently, period preservation is a very common phenomenon: it is not limited to the aforementioned period-doubling cascade but rather appears throughout a bifurcation diagram.

Periodic orbits anticipated by Theorem 1 and confirmed in Section \ref{sec:numerical} correspond to traveling waves in a one-dimensional HWCML and to periodic patterns in a multidimensional HWCML. Such patterns have been studied in homogeneous CMLs \cite{He1996,Silva2014}, and our results can help to describe such dynamics in heterogeneous CMLs both near and far from bifurcations. Our observation about period resilience implies that there will be many different patterns with the same period. Small changes in an HWCML can change the specific pattern, but the period itself is rather robust.

Our results also have implications in applications. A toy macroscopic traffic flow model, governed by the logistic map, was proposed in \cite{shih}. The derivation of the model is based on very general assumptions involving speed and density. When these assumptions are satisfied, one can use the model to help examine the evolution of flows of pedestrians, flows in a factory, and so on.  When such flows interact weakly, then equations of the form that we discussed in Section \ref{sub:PeriodicOrbits} can be useful for such applications. For example, one could do a simple examination of the temporal evolution of two groups of football fans around a stadium (or of sheep around an obstacle \cite{sheep}). The two groups have different properties, so suppose that they are governed by an HWCML. From our results, if each group is regularly entering the stadium on its own (i.e., their behavior is periodic), then both groups considered together would continue to enter regularly at the same rate, provided that the interaction between the two groups is weak. This suggests that it would be interesting to explore a security strategy that models erecting a light fence to ensure that the interaction between the two groups remains weak.

The model in Ref.~\cite{shih} also admits chaotic traffic patterns. One can construe the intermittent traffic flow in a traffic jam as being formed by regular motions (i.e., a laminar regime) and a series of acceleration and braking (i.e., chaotic bursts). Our results give the temporal evolution of such a laminar regime in a chaotic intermittent flow if the interaction between entities is weak (i.e., when the laminar regime is long, as we discussed in Section \ref{sub:TypeI}). Indeed, as has been demonstrated experimentally for the flow of sheep around an obstacle \cite{sheep}, it is possible to preserve laminar behavior for a longer time through the addition of an obstacle.

%%%%

\section*{Acknowledgements}

We are grateful to the anonymous referee for his/her enlightening and detailed suggestions. We also thank Daniel Rodr\'iguez P\'erez for his help in the preparation of this manuscript.

%%%%%%%%%%%%%%%%%%%% references %%%%%%%%%%%%%%

%%%%%%%%%%%%%%%%%%%%%%%%%%%%%%%%%%%%
% para que los flotantes salgan a partir de aqu????

\clearpage

\section*{Figure Legends}

%!!!

%!!!

%%%%%%%%%%%%%%%%%%%%%%%%%%%%%%%%%%%%

% para que los flotantes salgan a partir de aqu????

%%%%%%

\clearpage

\section*{Tables}

%%%%%%%%%%%%%%%%%%%%%%%%%%%%%%%%%%%%

\begin{table}[!hhh]
	\caption{\label{tab:TABLA1}
	         {\bf Period of the CML (\ref{equA}) for $r_2=1.9$ and $\varepsilon=0.001$}.
	         The parameter $r_1$ indicates when the logistic map, which is satisfied by the free oscillator $X(n)$, exhibits orbits of various periods during a period-doubling cascade in the window of period-3 orbits in the bifurcation diagram. Although the period of $X(n)$ changes, the period of the CML remains the same.
	}
\begin{tabular}{|l|l|l|}\hline
$r_1$ & Period of $X(n)$ &	Period of the CML\\ \hline
$3.831874$  & $3$           & $\lcm(3,4)=12$\\ \hline
$3.844568$  & $3 \times 2$  & $\lcm(3 \times 2,4)=12$ \\ \hline
$3.848344$  & $3 \times 2^2$& $\lcm(3 \times 2^2,4)=12$\\ \hline
\end{tabular}
\end{table}
%%%%%%%%%%%%%%%%%%%%%%%%%%%%%%%%%%%%

\end{document}